\documentclass[prx,aps,twocolumn,showpacs,superscriptaddress,nofootinbib]{revtex4-1}
\usepackage[colorlinks=true,citecolor=blue,linkcolor=magenta]{hyperref}
\usepackage{graphicx}
\usepackage{dcolumn}
\usepackage{bm}
\usepackage[bbgreekl]{mathbbol}
\usepackage{amsmath,dsfont}
\usepackage[normalem]{ulem}
\usepackage{color}
\usepackage{booktabs}

\newcommand{\Cth}{C_\text{th}}

\newcommand{\xtil}{\tilde{x}}
\newcommand{\ttil}{\tilde{t}}
\newcommand{\erf}{\text{erf}}

\newcommand{\taubar}{\bar{\tau}}

\begin{document}
\title{Diffusive wave dynamics beyond the continuum limit}

\author{Paul B. Dieterle}
\address{Department of Physics, Harvard University, Cambridge, MA 02138, USA}

\author{Ariel Amir}
\email{arielamir@seas.harvard.edu}
\address{John A. Paulson School of Engineering and Applied Sciences, Harvard University, Cambridge, MA 02138, USA}

\begin{abstract}
Scientists have observed and studied diffusive waves in contexts as disparate as population genetics and cell signaling. Often, these waves are propagated by discrete entities or agents, such as individual cells in the case of cell signaling. For a broad class of diffusive waves, we characterize the transition between the collective propagation of diffusive waves -- in which the wave speed is well-described by continuum theory -- and the propagation of diffusive waves by individual agents. We show that this transition depends heavily on the dimensionality of the system in which the wave propagates and that disordered systems yield dynamics largely consistent with lattice systems. In some system dimensionalities, the intuition that closely packed sources more accurately mimic a continuum can be grossly violated.
\end{abstract}

\maketitle

Partial differential equations (PDEs) pervade quantitative studies of biological phenomena. One famous success story of PDEs in biology is the study of diffusive waves, first begun by Luther in 1906 \cite{luther1906} and furthered by Fisher and Kolmogorov \textit{et al.} in 1937 \cite{fisher1937,Kolmogorov1937}. In the typical setting, a homogeneous environment can propagate a diffusive wave by picking up some diffusing element -- an organism or a signaling molecule -- and producing more of that same element -- through reproduction or secretion of the same entity. The process of diffusion and reproduction can be represented by a partial differential equation, the dynamics solved by way of a traveling wave ansatz, and the wave speed deduced analytically (in certain scenarios). This type of machinery has been adopted to understand traveling waves governing apoptosis \cite{cheng2018}, ecological range expansions \cite{Barton2011,tanaka2017,fisher1937,Kolmogorov1937,ghandi2016}, cell signaling \cite{noorbakhsh2015,kessler1993,Nolet2020}, mitosis \cite{chang2013}, and microbial consortia \cite{parkin2018,youk2014}.

However, in many of these applications, one is led to believe that the underlying PDE models hold in the case of discrete sources. For instance, cell signaling waves are propagated by discrete cells \cite{noorbakhsh2015,kessler1993}, mitotic and calcium waves in Xenopus are paced by discrete sources (nuclei and receptor channels, respectively) \cite{Nolet2020,Afanzar2020,Keener2000,Kupferman1997,Ponce1999,Mitkov1998}, and reproduction in mobile populations happens in centralized locations like cities. In such cases, it is not necessarily valid to assume a homogeneous environment when calculating the properties of the diffusive wave; one must instead solve the dynamics for a different set of PDEs -- ones with discrete sources.

Here, for a broad class of models, we show that the discreteness of source terms can drastically alter the diffusive wave dynamics. We find that in many cases, one can characterize these corrections analytically for both ordered and disordered systems. The corrections can be dramatic -- especially in the case of Fisher wave dynamics. In the extreme limit of very disparate sources, we find the Fisher wave speed is independent of the source density and the diffusion constant. By examining the effects of discreteness, we see a clear distinction between diffusive waves that can be driven by activation at or beyond the wave front (termed ``pulled waves" in the literature) and those that rely on concentration build up from sources behind the front (termed ``pushed waves"). In the extreme limit of sparse sources, we show that pulled waves are driven by self-amplification while pushed waves rely on excitation from a nearest neighbor. We find that the effects of discreteness on asymptotic wave propagation are dimension-dependent. For systems with diffusion in two dimensions, packing sources ever closer does not make the system behave more like a continuum; for diffusion in three dimensions, packing sources ever closer makes the system behave \textit{less} like a continuum. We also, for the first time, characterize the effects of disorder on diffusive wave propagation, finding that for many system dimensionalities a simple lattice model suffices for understanding the dynamics of diffusive waves in disordered systems.

\section*{Model construction}

To begin, we consider a model of diffusive wave propagation in which we monitor the concentration, $c$, of some elements -- e.g., signaling molecules in the case of cell signaling, organisms in the case of Fisher waves -- which diffuse with diffusion constant $D$. In addition to basic diffusion, the elements can be secreted by the background according to some concentration-dependent production function $F(c)$. Combining production with diffusion, we arrive at a generic model for the propagation of the concentration in time:

\begin{equation}
\label{eq:modelgeneric}
\frac{\partial c}{\partial t} = D\nabla^2c+F(c).
\end{equation}

Concretely, one could imagine a background of cells of density $\rho$ which secrete diffusible molecules at some rate, $a$, above a certain threshold concentration, $\Cth$; in this case, with $\Theta[.]$ the Heaviside step function, $F(c) = a\rho\Theta[c-\Cth]$ \cite{dieterle2020,kessler1993,mckean1970}. Inspired by ecology, Fisher and Kolmogorov \textit{et al.} considered production according to a logistic growth process in which $F(c) = ac(1-c)$ \cite{fisher1937,Kolmogorov1937}. Importantly, the dynamics driven by the Fisher-Kolmogorov production function depend only on the fact that $F(c) \sim c$ for $c\ll 1$. We will consider a class of functions that interpolate between the Fisher-Kolmogorov and Heaviside limits, in which the production function is of a Hill function form, $F(c) = a\rho\frac{c^n}{c^n+\Cth^n}$. For $n=1$, the production function scales linearly in $c$ for $c\ll\Cth$: $F(c\ll\Cth) \sim c$. In this limit, as we will show shortly, our model reproduces Fisher-Kolmogorov dynamics. For $n\rightarrow\infty$, $F(c) = a\rho\Theta[c-\Cth]$ and we will reproduce Heaviside dynamics. In total, then, we commit ourselves to studying equations of the form:

\begin{equation}
\label{eq:modelHill}
\frac{\partial c}{\partial t} = D\nabla^2c+a\rho\frac{c^n}{c^n+\Cth^n}
\end{equation}

\noindent along with the discrete analogs thereof.

Several prior works inspired by calcium waves \cite{Keener2000,Mitkov1998,Ponce1999,Kupferman1997} have studied the one dimensional discrete analogs of eq. \eqref{eq:modelgeneric}, accounting additionally for simple decay of $c$ by way of a term $-\gamma c$. These works cover a variety of production functions, primarily focusing on the form put forth in eq. \eqref{eq:modelHill} as $n\rightarrow \infty$. One such work treats discreteness as a perturbation to the continuum theory dynamics \cite{Keener2000}. We will instead opt for a solution that yields the wave speeds -- albeit through transcendental equations -- even far from the continuum limit. (Additional discrete corrections have famously been studied by Brunet and Derrida, among others, in the context of Fisher Waves \cite{brunet1997}.) Here, we will assume that decay is negligible and will seek to understand the corrections due to discreteness for all $n$. Moreover, whereas these works focused almost exclusively on the dynamics of lattice-like systems in one dimension, we will additionally characterize the dynamics under disorder and in all dimensions.

To take into account the effects of discreteness, we will assume that sources are point-like and thus that the discrete analogs of eq. \eqref{eq:modelHill} can be written using Dirac delta functions $\delta(.)$. Thusly, and by replacing the density $\rho$ with a sum over the locations $\mathbf{r}_j$ of all sources, we have

\begin{equation}
\label{eq:modelHillDiscrete}
\frac{\partial c}{\partial t} = D\nabla^2c+a\frac{c^n}{c^n+\Cth^n}\sum_j\delta(\mathbf{r}-\mathbf{r}_j).
\end{equation}

\noindent The goal of this work will be to study the relationship between the dynamics of eq. \eqref{eq:modelHill} and those of eq. \eqref{eq:modelHillDiscrete}. To start, we will review the standard, continuum theory wave dynamics of eq. \eqref{eq:modelHill} in all dimensions; we will then calculate the one-dimensional discrete model corrections to the continuum theory by considering eq. \eqref{eq:modelHillDiscrete} for all $n$ and with different distributions of $\mathbf{r}_j$; finally, specializing to the case of $n\rightarrow\infty$, we will solve for the dynamics in arbitrary dimensions, showing in the process that the nature of the wave speed corrections heavily depend on the dimensionality of the diffusive environment and the distribution of sources within it.

\section*{Continuum model solutions in all dimensions}

Let us begin our study of the continuum model of eq. \eqref{eq:modelHill} by considering a purely one-dimensional system, in which case we have

\begin{equation}
\label{eq:modelHill1D}
\frac{\partial c}{\partial t} = D\frac{\partial^2c}{\partial x^2}+a\rho\frac{c^n}{c^n+\Cth^n}
\end{equation}

\noindent with $\rho$ a density of cells in one dimension. By making a traveling wave ansatz $c(x,t) = c(\xtil = x-vt)$, in which a concentration wave travels with constant speed $v$ and $\xtil$ is the distance with respect to the wave front, one can eliminate the time derivative of eq. \eqref{eq:modelHill1D} and work instead with non-linear ordinary differential equations (ODEs) of the form

\begin{equation}
\label{eq:modelHill1DODE}
0 = v\frac{\partial c}{\partial\xtil}+D\frac{\partial^2c}{\partial \xtil^2}+a\rho\frac{c^n}{c^n+\Cth^n}.
\end{equation}

\noindent Our goal is to find the wave speed $v$ as a function of the physiological parameters $n$, $D$, $\Cth$, $a$, and $\rho$.

As $n\rightarrow 1$ or $\infty$, this task has a well-known solution. For the latter, one may substitute $c^n/(c^n+\Cth^n)\rightarrow\Theta[c-\Cth]$; recognize that $\xtil=0$ is the point at which $c(\xtil) = \Cth$; and stitch together solutions for $\xtil>0$ and $\xtil<0$, in which $\Theta[c-\Cth] = 1$ and $0$, respectively \cite{dieterle2020,kessler1993,Kupferman1997,mckean1970}. Throughout this paper, we will refer to the wave speed with sources distributed in $N$ dimensions, with diffusion in $M$ dimensions, and with Hill coefficient $n$ as $v_{N,M,n}$; the calculation described above reveals that

\begin{equation}
\label{eq:v11Inf}
v_{1,1,n\rightarrow\infty} = \sqrt{a\rho D/\Cth}.
\end{equation}

The solution for $n=1$ is more involved and appeals to mathematics first worked out by Fisher and Kolmogorov \textit{et al.} \cite{fisher1937,Kolmogorov1937}; here, one examines the region well beyond the wave front ($\xtil\gg D/v$), in which $c\ll \Cth$ and $c/(c+\Cth)\approx c/\Cth$. In these limits, we may consider a modified form of eq. \eqref{eq:modelHill1DODE}:

\begin{equation}
\label{eq:modelHill1Dn1}
0 \approx v\frac{\partial c}{\partial\xtil}+D\frac{\partial^2c}{\partial \xtil^2}+a\rho c/\Cth.
\end{equation}

\noindent One then plugs an ansatz of the form $c = c_0e^{-\gamma\xtil}$ into eq. \eqref{eq:modelHill1Dn1}, which results in a relationship between $v$ and $\gamma$:

\begin{equation}
\label{eq:wavespeedn1}
v = D\gamma+a\rho/\Cth\gamma.
\end{equation}

\noindent Given the restriction that $v,\gamma>0$, eq. \eqref{eq:wavespeedn1} appears to allow a range of solutions $v\geq 2\sqrt{a\rho D/\Cth}$; famously \cite{Kolmogorov1937,fisher1937}, the minimum speed allowed by eq. \eqref{eq:wavespeedn1} is the one that the wave attains, a fact that has been proven to hold with periodic and homogeneous production functions alike \cite{Cao2019}. Thus,

\begin{equation}
\label{eq:v111}
v_{1,1,1} = 2\sqrt{a\rho D/\Cth}.
\end{equation}

It is not a coincidence that $v_{1,1,1}$ and $v_{1,1,n\rightarrow\infty}$ both scale as $\sqrt{a\rho D/\Cth}$. As has been shown previously \cite{dieterle2020,luther1906}, this scaling can be deduced by dimensional analysis considerations. In brief, the source terms on the R.H.S. of eq. \eqref{eq:modelHill1D} are proportional to $a\rho$, meaning all concentration scales are also proportional to this factor. Thusly, the four parameters of $a$, $\rho$, $\Cth$, and $D$ are boiled down to two parameters, $D$ and $\Cth/a\rho$ with respective dimensions m$^2$/s and s. As we wish to calculate an emergent wave speed $v$ with units m/s, the only viable formula for $v$ is a scaling law of the form $v\sim \sqrt{a\rho D/\Cth}$.

While these results plainly hold in systems of dimensionality $(N,M) = (1,1)$, they also hold asymptotically in all systems of dimensionality $N=M$, so long as the dynamics are observed at radii $r\gg ND/v$ \cite{dieterle2020,tanaka2017}.

In contrast to systems with $N=M$, one could consider a system in which some $N$-dimensional distribution of sources is embedded within a higher-dimensional, $M>N$ diffusive environment. One example is an assemblage of cells sitting upon a two-dimensional plane with a semi-infinite extracellular medium above \cite{dieterle2020}. In such a scenario, assuming a semi-infinite diffusive environment with $N=1$ and $M=2$, eq. \eqref{eq:modelHill} becomes

\begin{equation}
\label{eq:modelHill1D2D}
\frac{\partial c}{\partial t} = D\left(\frac{\partial^2c}{\partial x^2}+\frac{\partial^2c}{\partial z^2}\right)+2a\rho\delta(z)\frac{c^n}{c^n+\Cth^n}
\end{equation}

\noindent with $x$ the dimension in which the sources live, $z$ the perpendicular dimension, and $\rho$ a one-dimensional density of sources. Within the aforementioned traveling wave ansatz, one can reduce this to a spatial PDE of the form

\begin{equation}
\label{eq:modelHill1D2DSpatialPDE}
0 = v\frac{\partial c}{\partial\xtil}+D\left(\frac{\partial^2c}{\partial \xtil^2}+\frac{\partial^2c}{\partial z^2}\right)+2a\rho\delta(z)\frac{c^n}{c^n+\Cth^n}.
\end{equation}

\noindent Before noting the exact solutions of the wave speed, we again appeal to dimensional analysis. As in the case of $N=M$ discussed above, the source terms on the R.H.S. of eq. \eqref{eq:modelHill1D} are proportional to $a\rho$ and thus we are again left with only two effective model parameters, $D$ and $\Cth/a\rho$, with which to construct a wave speed. However, unlike the case of $N=M$, in which $\Cth/a\rho\sim$s, we now have $\Cth/a\rho\sim$s/m. Thus, the only formula involving $D$ and $\Cth/a\rho$ which forms a valid wave speed, $v$, of units m/s is the $D$-independent scaling law of $v\sim a\rho/\Cth$.

For $n\rightarrow\infty$, one can solve eq. \eqref{eq:modelHill1D2DSpatialPDE} using a partial Fourier transform over $z$ and the methodology employed on eq. \eqref{eq:modelHill1DODE} to yield

\begin{equation}
\label{eq:v12Inf}
v_{1,2,n\rightarrow\infty} = 2a\rho/\pi\Cth.
\end{equation}

\noindent A similar approach with $n=1$ requires an unwieldy partial Fourier transform over $z$ of

\begin{equation}
\label{eq:modelHill1D2DODE}
0 = v\frac{\partial c}{\partial\xtil}+D\left(\frac{\partial^2c}{\partial \xtil^2}+\frac{\partial^2c}{\partial z^2}\right)+2a\rho\delta(z)c(\xtil,z)/\Cth.
\end{equation}

\noindent In Appendix D we show it is possible to solve for the dynamics of eq. \eqref{eq:modelHill1D2DODE} with Green's function methods. Doing so, we show that

\begin{equation}
\label{eq:v121}
v_{1,2,1} = 2a\rho/\Cth.
\end{equation}

\noindent In closing this section we note that, as in the case of $N=M$, eq. \eqref{eq:v12Inf} and eq. \eqref{eq:v121} hold for all systems with $N=M-1$ so long as the curvature of the wave front is negligible (in a spherically symmetric system, this requirement is met when observing the dynamics at radii $r\gg ND/v$). We also note that systems for which $N = M-j$ and $j\geq 2$ lack wave-like solutions in the continuum. Here, dimensional analysis reveals a wave speed relationship such that $v$ increases ($j>2$) or remains constant ($j=2$) with increasing $\Cth$. The issue is that embedding a continuum of diffusive sources in a higher-dimensional space creates a divergent concentration at the sources when the discrepancy between the source dimension and the diffusive dimension is more than one.

The results for continuum systems with $N=M$ or $N=M-1$ are summarized in Table 1.

{
\begin{table}
\centering
\begin{ruledtabular}
\begin{tabular}{c|c|c}
 & $N=M$ & $N=M-1$ \\
\hline
$n = 1$ & $v_{N,N,1} = 2\sqrt{a\rho D/\Cth}$ & $v_{N,N+1,1} = 2a\rho/\Cth$ \\
\hline
$n\rightarrow\infty$ & $v_{N,N,n\rightarrow\infty} = \sqrt{a\rho D/\Cth}$ & $v_{N,N+1,n\rightarrow\infty} = 2a\rho/\pi\Cth$ \\
\end{tabular}
\end{ruledtabular}

\caption{\textbf{Summary of continuum theory wave speeds:} Here, we summarize the diffusive wave speed for systems with sources in $N$ dimensions and diffusion in $M$ dimensions. We consider system dimensionalities of $N=M$ and $N=M-1$ in the limits $n = 1$ and $n\rightarrow\infty$. As shown in the main text, dimensional analysis demands that systems with $N=M-1$ display a diffusion constant ($D$) independent wave speed and scale differently in the source density $\rho$, emission rate $a$, and threshold concentration $\Cth$.}
\label{table:dynamicssummary}
\end{table}
}

\section*{Discrete model solutions in one dimension}

We now seek to depart from the continuum theory of diffusive waves by focusing on waves propagated by discrete sources. To do so, we examine the one-dimensional version of the discrete model put forth in eq. \eqref{eq:modelHillDiscrete}:

\begin{equation}
\label{eq:modelHillDiscrete1D}
\frac{\partial c}{\partial t} = D\frac{\partial^2c}{\partial x^2}+a\frac{c^n}{c^n+\Cth^n}\sum_j\delta(x-x_j).
\end{equation}

\noindent Here, we have discrete sources at positions $x_j$, each of which responds to the local concentration $c$ by emitting diffusible molecules at some rate $ac^n/(c^n+\Cth^n)$. In this setting, we would like to find something akin to the wave speed calculated in the continuum limit.

To begin, let us consider the dynamics of sources on a lattice with separation $d$ (i.e., density $\rho = 1/d$) as in Fig. 1A/B. We anticipate that the wave speed will be close to that of the continuum model for sufficiently small $d$ -- but how small?

Examining the $n=1$ solution above, we see that our ansatz has a beyond-the-front solution of $e^{-vx/2D}$. Thus, $D/v$ is roughly the length scale of the diffusive wave, and we expect that when many sources fit inside this length scale -- i.e., when $dv/D\ll 1$ -- the continuum and discrete models will agree. Of course, $v$ itself depends upon $d$. Within the continuum theory $\rho = 1/d\implies v\sim \sqrt{aD/d\Cth}$, which implies that $d\ll D\Cth/a$ is the limit in which the two models should agree and that $D\Cth/a$ is the length scale against which we should compare $d$.

We can come about this length scale through simpler means of dimensional analysis. Omitting $d$, the only other parameters in our model are $\Cth$, $a$, and $D$. As in the continuum limit, we may eliminate $a$ by noting that the source terms in eq. \eqref{eq:modelHillDiscrete1D} are all proportional to $a$. Thus, we are left with $D$ (units of m$^2$/s) and $\Cth/a$ (units of $s/m$) as the only other model parameters. Using these parameters to construct a length scale, we see there is a unique length scale $D\Cth/a$ against which we may compare $d$. In essence, when $\Cth/a$ or $D$ is sufficiently small, only the nearest neighbors contribute to the local concentration and we can expect deviations from the continuum theory. Keener colorfully refers to this limit as the ``saltatory" regime, in which the wave front hops form one source to the next \cite{Keener2000}. With this length scale in mind, let us conduct a rigorous analysis for $n\rightarrow \infty$ and $n=1$.

\subsection*{Lattice solution for \texorpdfstring{$n\rightarrow\infty$}{hi}}

In the limit $n\rightarrow\infty$, it is possible to construct a relationship which gives a unique wave speed $v$ as a function of $d$, $D$ and $\Cth/a$. To do so, we consider a lattice of point sources, as in Fig. 1B, through which a diffusive wave has been traveling at speed $v$. We set the time and space origins such that the concentration at $t, \xtil = 0$ is $\Cth$; doing so, we stipulate that the source a distance $jd$ away has been continuously emitting at a rate $a$ since time $t=-jd/v$. Using the diffusion equation Green's function for a point-like source, we can therefore see that the concentration, $c_j$, created by the source a distance $jd$ away from the origin is given by

\begin{multline}
\label{eq:cjInf}
c_j = a\int_{-jd/v}^0d\ttil~e^{j^2d^2/4D\ttil}/\sqrt{-4\pi D\ttil} = \\
\frac{jad}{2D}\left(e^{-jdv/4D}\sqrt{\frac{4D}{\pi jdv}}+\erf{\sqrt{\frac{jdv}{4D}}}-1\right)
\end{multline}

\noindent while the threshold concentration, $\Cth$, obeys a self-consistency relationship given by summing over all the sources:

\begin{equation}
\label{eq:latticeCthInf}
\Cth = \sum_{j=1}^\infty c_j.
\end{equation}

\noindent In constructing eq. \eqref{eq:latticeCthInf}, we have successfully derived a relationship between $v$ and the system parameters $\Cth/a$, $d$, and $D$. We will now explore this relationship to test the intuition we developed through dimensional analysis.

By examination of eq. \eqref{eq:cjInf}, we plainly see that the length scale with which to compare $d$ is $D/v$. The latter is the length scale of the diffusive wave, as mentioned above. Let us assume that $d\ll D/v$ -- i.e., that many sources fit within the length scale of the diffusive wave -- in which case one may turn eq. \eqref{eq:latticeCthInf}'s sum over $j$ into an integral and deduce that

\begin{multline}
\label{eq:latticeCthInfInt}
d\ll D/v:~\Cth \approx \int_0^\infty dj~c_j = aD/v^2d \implies \\
v = \sqrt{aD/d\Cth},
\end{multline}

\noindent which is exactly the continuum theory result of eq. \eqref{eq:v11Inf} with $\rho = 1/d$. Using our expression for $v$, we may deduce that the assumption of $d\ll D/v$ is equivalent to:

\begin{equation}
\label{eq:lengthScaleComp}
d\ll D/v = D\sqrt{d\Cth/aD}\implies d\ll \Cth D/a,
\end{equation}

\noindent exactly as predicted by our dimensional analysis. In this limit, the concentration wave can be understood as a collective phenomenon in which many sources contribute to the concentration at the wave front (Fig. 1F) and the wave speed is approximately that of our continuum theory (Fig. 1C). Thus, when the dimensionless parameter $\phi = ad/D\Cth$ is small, the collectivity is high and \textit{vice versa}.

In the contrasting limit of $d\gg \Cth D/a$, the concentration at the wave front is primarily generated by the nearest source (Fig. 1F) and the wave speed is considerably less than that predicted by the continuum theory. We will return to this limit shortly.

\begin{figure*}[t!]
    \centering
    \includegraphics[width=17.2cm]{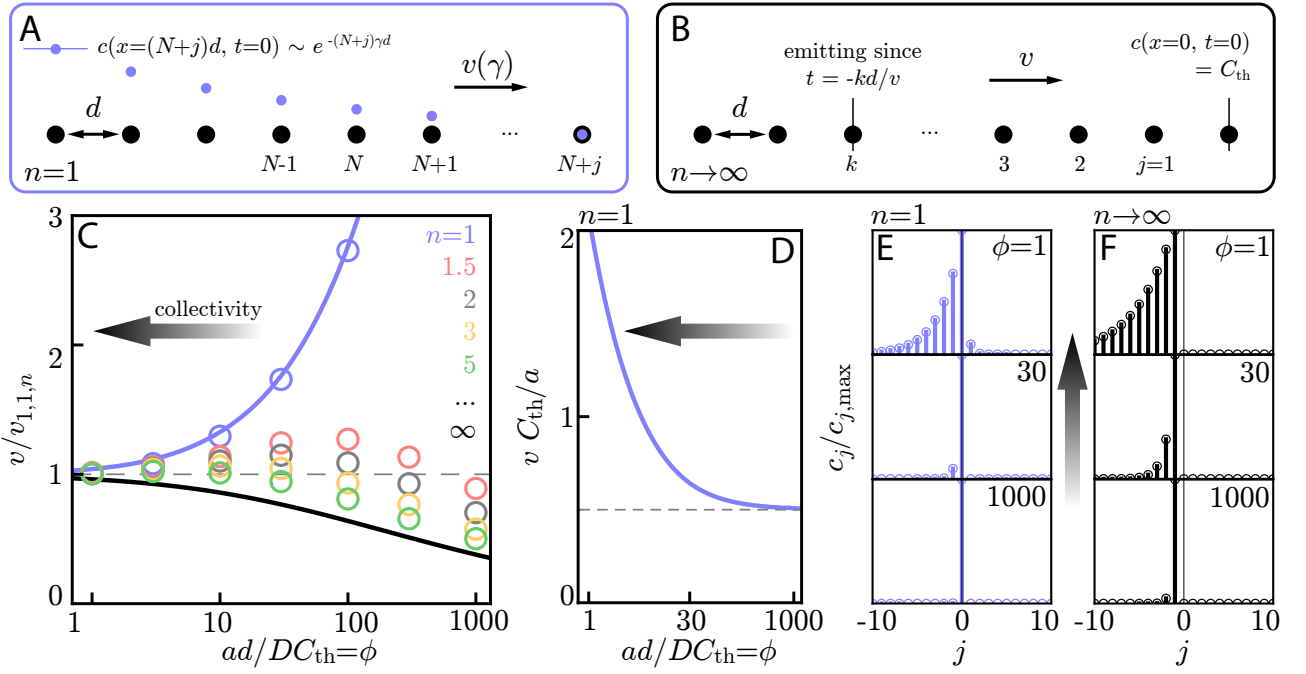}
    \caption{\textbf{Discrete lattice wave dynamics in 1D.} \textbf{A:} Schematic demonstrating our solution method for Fisher-like waves with $n=1$. Here, we examine the region well beyond the wave front and make an ansatz that the concentration at source $N+j$ is given by $c(x=(N+j)d,t=0)\sim e^{-(N+j)\gamma d}$. This ansatz allows a range of solutions $v(\gamma)$, the minimum of which is the realized wave speed. \textbf{B:} Schematic demonstrating the solution method for Heaviside waves with $n\rightarrow\infty$. Here, the concentration at the wave front is given by $\Cth$ while the source a distance $kd$ from the wave front has been emitting since time $t = -kd/v$. \textbf{C:} Discrete wave speeds divided by continuum wave speed as a function of $ad/D\Cth = \phi$ for the $n=1$ theory (blue line), $n\rightarrow\infty$ theory (black line), and finite $n$ numerics (colored circles). For all $n>1$, large separation between sources leads to a sub-continuum wave speed while for $n=1$ the wave speed increases compared with its continuum value. At small values of $\phi$, the wave can be understood as a collective effect that is well-described by a continuum theory. \textbf{D:} Theoretical wave speed for $n=1$ as stipulated by eq. \eqref{eq:latticeCthn1}. At large values of $\phi$, the wave speed attains a $d$- and $D$-independent value of $a/2\Cth$. \textbf{E:} Concentration at source $N$ generated by source $N+j$ for various values of $\phi$ and for $n=1$. At small $\phi$, many sources contribute to the concentration at source $N$ while as $\phi$ increases, the wave propagation becomes a few-body phenomenon. For $\phi=1000$, we see the extreme limit of a pulled wave, in which the concentration at the wave front is almost entirely due to self-amplification.} \textbf{F:} Concentration at the wave front generated by source $j$ as $n\rightarrow\infty$. As with the case of $n=1$, many sources contribute to the concentration at the wave front when $\phi$ is small. As $\phi$ increases, the wave propagation becomes a few-body problem. In contrast to the case of $n=1$, for $\phi=1000$ the wave front is pushed by the nearest neighbor behind the front.
    \label{fig:1}
\end{figure*}

\subsection*{Lattice solution for \texorpdfstring{$n=1$}{hi} and numerics for all \texorpdfstring{$n$}{hi}}

Next, we consider the $n=1$ limit of eq. \eqref{eq:modelHillDiscrete1D}. To address this limit, we again consider a lattice of point sources, as in Fig. 1A, in which a diffusive wave with speed $v$ has been traveling. Examining the region of space well beyond the wave front in which $c\ll \Cth$, we follow the example of Fisher \cite{fisher1937} and propose an ansatz that the concentration some distance $\xtil = Nd$ beyond the wave front is given by:

\begin{equation}
\label{eq:discrete1Dn1Ansatz}
c(\xtil = Nd, t) = c_0e^{-\gamma Nd+\gamma vt}.
\end{equation}

\noindent In the limit of $c\ll \Cth$, the production function may be approximated as $c/(c+\Cth)\approx c/\Cth$. Thus, the concentration, $c_j$, generated by the source $\xtil = (N+j)d$ as measured at $\xtil = Nd$ and $t=0$ can again be calculated by integrating the production function of source $N+j$ against the Green's function for the diffusion equation:

\begin{multline}
\label{eq:cj1Dn1}
c_j = \frac{ac_0e^{-\gamma Nd}}{\Cth}\int_{-\infty}^0 dt~\frac{e^{j^2d^2/4Dt}e^{-\gamma jd+\gamma vt}}{\sqrt{-4\pi Dt}} = \\
\frac{ac_0e^{-\gamma Nd}}{2\Cth\sqrt{\gamma Dv}}e^{-d\left(j\gamma+|j|\sqrt{\gamma v/D}\right)}.
\end{multline}

\noindent One may construct a self-consistency relationship along the lines of eq. \eqref{eq:latticeCthInf} by recognizing that the concentration $c(\xtil = Nd, t=0)$ is merely the sum of all the $c_j$, so that

\begin{equation}
\label{eq:latticeSelfConsist}
c(\xtil = Nd, t=0) = \sum_{j=-\infty}^\infty c_j,
\end{equation}

\noindent which can be written in full as

\begin{multline}
\label{eq:latticeCthn1}
\frac{2\Cth\sqrt{D\gamma v}}{a} = \sum_{j=-\infty}^\infty \exp\left[-j\gamma d-|j|d\sqrt{\gamma v/D}\right] = \\
1+\frac{1}{e^{d\gamma+d\sqrt{\gamma v/D}}-1}+\frac{1}{e^{-d\gamma+d\sqrt{\gamma v/D}}-1}.
\end{multline}

\noindent Through eq. \eqref{eq:latticeCthn1}, we are endowed with a relationship, like eq. \eqref{eq:latticeCthInf}, that fully specifies $v$ given the physiological parameters $\Cth/a$, $d$, and $D$ -- but with one exception. As in the famous case of Fisher's equation, we must also specify $\gamma$; and, as in the case of Fisher's equation, the value of $\gamma$ that is selected is the one that minimizes $v$. Given eq. \eqref{eq:latticeCthn1}, one can find the wave speed $v$ by minimizing $v$ over all $\gamma$ subject to the transcendental relationship of eq. \eqref{eq:latticeCthn1}. That the wave speed achieved is the minimum has been proven for all periodic systems in beautiful recent work \cite{Cao2019}.

As in the case of $n\rightarrow\infty$, our goal here is to see how the wave speed stipulated through the above calculations relates to the continuum theory of eq. \eqref{eq:v111}. Appealing to our previous intuition, we can guess that $d\ll \Cth D/a$ will yield a wave speed consistent with the continuum theory. Indeed, we may expand eq. \eqref{eq:latticeCthn1} to lowest order in $d$ for $d\ll \gamma d,~d\sqrt{\gamma v/D}$; doing so leaves one with exactly the relationship of eq. \eqref{eq:wavespeedn1} with $\rho = 1/d$. The value of $\gamma$ which minimizes $v$ is $\gamma = \sqrt{d\Cth D/a}$, meaning that our imposition of $d\ll \gamma d,~d\sqrt{\gamma v/D}$ is equivalent to $d\ll \Cth D/a$. In this limit, the wave speed $v$ agrees well with the continuum theory (Fig. 1C) and the effect is a collective one in which many sources contribute to the concentration generated at every other source (Fig. 1E).

In the opposite limit of $d\gg \Cth D/a$, the primary driver of the concentration generated at each source is the concentration generated by the source itself (Fig. 1E). The source is seeded with some small concentration from its nearest neighbor, then self-amplifies; thus, unlike in the case of $n\rightarrow\infty$ in which only the nearest neighbor matters, we now must account for both the nearest neighbor and the self-interaction. Surprisingly, the wave speed in this limit is independent of $d$ and $D$, as we will now show and as is pictured in Fig. 1D. In this limit, one can approximate the sum in eq. \eqref{eq:latticeCthn1} by considering only the dominant $j=0,-1$ terms (self and nearest-neighbor). Doing so yields

\begin{equation}
\label{eq:latticeCthn1dLarge}
d\gg \Cth D/a:~\frac{2\Cth\sqrt{D\gamma v}}{a} = 1+e^{\gamma d-d\sqrt{\gamma v/D}}.
\end{equation}

\noindent Treating $v$ as a function $v(\gamma)$, one can take a derivative of both sides of the above equation with respect to $\gamma$. Rearranging terms, we can see that $\partial v/\partial \gamma = 0$ in the $d\rightarrow\infty$ limit requires $\gamma-\sqrt{\gamma v/D}\sim\log d/d$ and thus that

\begin{multline}
\label{eq:gammaMindLarge}
d\gg \Cth D/a:~\gamma = v/D \implies v = a/2\Cth,\\
\gamma = a/2D\Cth
\end{multline}

\noindent where the relationship after the arrow is had by plugging the relationship before the arrow into eq. \eqref{eq:latticeCthn1dLarge}.

It is strange that, in the $d\gg \Cth D/a$ limit, a wave propagated between discrete sources and driven by diffusion has a speed that depends neither on the distance between the sources nor on the rate of diffusion. To understand how this can be, we return to the ansatz of eq. \eqref{eq:discrete1Dn1Ansatz} and plug in the values of eq. \eqref{eq:gammaMindLarge}, which yields

\begin{equation}
\label{eq:dLargeAnsatz}
d\gg \Cth D/a:~c(\xtil=Nd,t) = c_0e^{-\frac{Nda}{2D\Cth}+\frac{a^2t}{4D\Cth^2}}.
\end{equation}

An examination of eq. \eqref{eq:dLargeAnsatz} shows that the wave speed is indeed $d$- and $D$-independent in the large-$d$ limit. We see that $c(\xtil = Nd,t) = c\left[\xtil = (N+1)d, t+2\Cth d/a\right]$, meaning that the concentration wave travels a distance $d$ in a time $\tau = 2\Cth d/a$. As a result, the wave speed $v = d/\tau$ is independent of $d$. In essence, increasing $d$ results in an exponentially smaller initial concentration at any point source but gives that point source more time to self-amplify -- effects that exactly cancel when calculating the wave speed.

Similarly, increasing the diffusion constant, $D$, results in an exponential increase in the initial concentration at a point source (see the factor of $e^{-Nda/2D\Cth}$ in eq. \eqref{eq:dLargeAnsatz}) due to the fact that molecules can more quickly diffuse from source to source; however, this effect is exactly cancelled by the decrease in the concentration's growth rate, $a^2/4D\Cth^2$, and results in the $D$-independent hopping time between sources, $\tau = 2\Cth d/a$, calculated above.

To understand the origin of this $d$- and $D$-independence, we must therefore understand why the concentration takes the precise form of eq. \eqref{eq:dLargeAnsatz}. To do so, let us consider a single source seeded with some small concentration $c_0\ll \Cth$ at $t=0$ and secreting with a concentration-dependent rate $ac/\Cth$. In this scenario, the concentration, $c$, at the source at time $t$ is given self-consistently given by integrating $ac/\Cth$ against the diffusion equation Green's function, thus giving us

\begin{equation}
\label{eq:cSingleSource}
c(t) = \frac{a}{\Cth\sqrt{4\pi D}}\int_0^td\ttil~c(\ttil)/\sqrt{t-\ttil},
\end{equation}

\noindent which can be solved with an ansatz $c_0e^{\Gamma t}$ for $t\gg 1/\Gamma$ by sending $t\rightarrow\infty$ in the integral bound, telling us that the concentration at the source indeed grows exponentially with rate

\begin{equation}
\label{eq:gammaSingleSource}
\Gamma = a^2/4D\Cth^2.
\end{equation}

\noindent Comparing this exponential growth rate of the concentration at a single point source with the traveling wave dynamics stipulated by eq. \eqref{eq:dLargeAnsatz}, we can plainly see that the growth rate of the concentration at sources well beyond the wave front is exactly the exponential growth rate of the concentration at a single, self-amplifying point source.

Moreover, by integrating the Green's function against the concentration production rate of $ac/\Cth$ for $c\ll\Cth$, we can deduce that a self-amplifying point source that has been growing for a time $t\gg\Gamma,~r^2/D$ creates a concentration profile given by

\begin{multline}
\label{eq:cSingleSourceProfile}
c(x,t) = \frac{a}{\Cth\sqrt{4\pi D}}\int_0^td\ttil~\frac{e^{-x^2/4D(t-\ttil)}e^{\Gamma\ttil}}{\sqrt{t-\ttil}}\sim\\
\int_0^\infty d\ttil~\frac{e^{-x^2/4D(t-\ttil)}e^{\Gamma\ttil}}{\sqrt{t-\ttil}}\sim e^{\Gamma t-x\sqrt{\Gamma/D}}.
\end{multline}

\noindent This is precisely the profile that the diffusive wave generates in the limit $d\gg\Cth D/a$, see eq. \eqref{eq:dLargeAnsatz}. Thus, we can understand the temporal dependence of eq. \eqref{eq:dLargeAnsatz} as the self-amplification rate of a single source and the spatial dependence of eq. \eqref{eq:dLargeAnsatz} as the the concentration profile that results from this self-amplification.

\subsection*{Transition to pushed waves for \texorpdfstring{$n>1$}{hi}}

While in the case of $n=1$ we may solve for the dynamics through a Green's function analysis of eq. \eqref{eq:modelHillDiscrete1D} in the $c\ll \Cth$ limit, the same analysis does not permit solutions for $n>1$ -- not even for $n = 1.0001$. In fact, as we now show, the dynamics for $n>1$ are fundamentally different than the dynamics of $n=1$, as is well-known from the studies of continuum models like eq. \eqref{eq:modelHill1D}, in which $n=1$ waves are termed ``pulled" by self-amplification in the $c\ll \Cth$ limit while $n>1$ waves are ``pushed" by the concentration generated sources behind and around the wave front \cite{deneke2018,Barton2011,tanaka2017,ghandi2016}.

To see this, we start by calculating the self-amplification dynamics of a point source with $n>1$ in the limit of $c\ll \Cth$. In this limit, $c^n/(c^n+\Cth^n)\approx c^n/\Cth^n$ and we may construct a self-consistency relationship of the form eq. \eqref{eq:cSingleSource} by integrating the emission rate $ac^n/\Cth^n$ against the Green's function for the diffusion equation and setting that equal to the local concentration at some later time $t$:

\begin{equation}
\label{eq:cSingleSourcen}
c(t) = \frac{a}{\Cth^n\sqrt{4\pi D}}\int_{-\infty}^td\ttil~c^n(\ttil)/\sqrt{t-\ttil},
\end{equation}

\noindent Fixing the time origin such that $c(t=0) = c_0$, the concentration grows as (with $\alpha \sim -a^2/4\Cth^{2n}D$, see Appendix A for full expression)

\begin{equation}
\label{eq:singleSourcenGrowth}
c(t) = \left[\alpha t+c_0^{-2(n-1)}\right]^{-\frac{1}{2(n-1)}},
\end{equation}

\noindent meaning that, for $c_0\ll\Cth$, the time for $c(t)$ to double scales as

\begin{equation}
\label{eq:singleSourcenTime}
t \sim (\Cth/c_0)^{2(n-1)}.
\end{equation}

\noindent In contrast, for any $n$, a diffusive wave (whether propagated by discrete or continuous sources, see Appendix B) will create a concentration profile in the limit $c\ll\Cth$ that decays exponentially in distance beyond the wave front and thus grows exponentially in time. Therefore, the time scale needed for the wave to pass through a distant point source with initial concentration $c_0$ is roughly given by:

\begin{equation}
\label{eq:diffWavePropTime}
t\sim \log\left[\Cth/c_0\right].
\end{equation}

\noindent We can therefore see that for $n>1$, individual point-sources cannot propagate an initial concentration quickly enough to exceed $\Cth$ by the time the wave passes through; thus, the concentration at sources well beyond the wave front (where $c\ll\Cth$) is primarily generated not through the self-interaction of sources in that region but rather through the build up of concentration generated by sources behind or around the wave front. This is precisely the phenomenon of a ``pushed" wave. This distinction is shown clearly in Figs. 1E/F. For large $ad/D\Cth$, we can see that the concentration at the wave front for $n=1$ is generated almost entirely by self-amplification. In contrast, the concentration at the wave front for $n\rightarrow\infty$ is generated almost solely by the source immediately behind the wave front.

Because the wave front for systems with $n>1$ is pushed by the concentration generated by sources behind the front, we expect that for sufficiently large separation -- in the limit where only nearest neighbors play a role -- the wave speed will behave similarly to that of a system with $n\rightarrow\infty$ in that the wave front will hop to the source beyond the wave front primarily as a result of concentration build up generated by the source immediately inside the wave front. This is a necessity because sources with $n>1$ in the exponential tail of the diffusive wave cannot excite themselves quickly (eq. \eqref{eq:singleSourcenTime}), but instead rely on the neighbors to ``send them" the diffusive wave.

A necessary consequence of this fact is that the wave speed will fall in comparison to the continuum wave speed. This is because diffusion requires a time $\tau\sim d^2$ to bring a neighboring source over the threshold concentration. The apparent wave speed of this process is then $v\sim d/\tau\sim 1/d$. In comparison, the continuum theory wave speed scales as $v_{1,1,n}\sim 1/\sqrt{d}$. Thus, for $n>1$ and $ad/D\Cth\gg 1$, the discrete source wave speed will fall below the continuum theory wave speed as $ad/D\Cth\rightarrow\infty$.

To affirm this result, we turned to numerical simulation. In Fig. 1C, we plot the wave speeds for various values of $n$ and $ad/D\Cth$. We obtained these values through numerical simulation of eq. \eqref{eq:modelHillDiscrete1D} and compared them to their continuum theory analog through numerically solving eq. \eqref{eq:modelHill1D} (see Appendix C for numerical details). As expected, systems for all values of $n$ agree well with their continuum theory analogs for small $ad/D\Cth$. For large $ad/D\Cth$, systems with $n=1$ display a super-continuum wave speed; in contrast, for sufficiently large $ad/D\Cth$, systems with $n>1$ display a sub-continuum wave speed as they rely not on self-excitation but on the comparatively slower process of diffusion between discrete sources around the front.

\begin{figure}[t!]
    \centering
    \includegraphics[width=8.6cm]{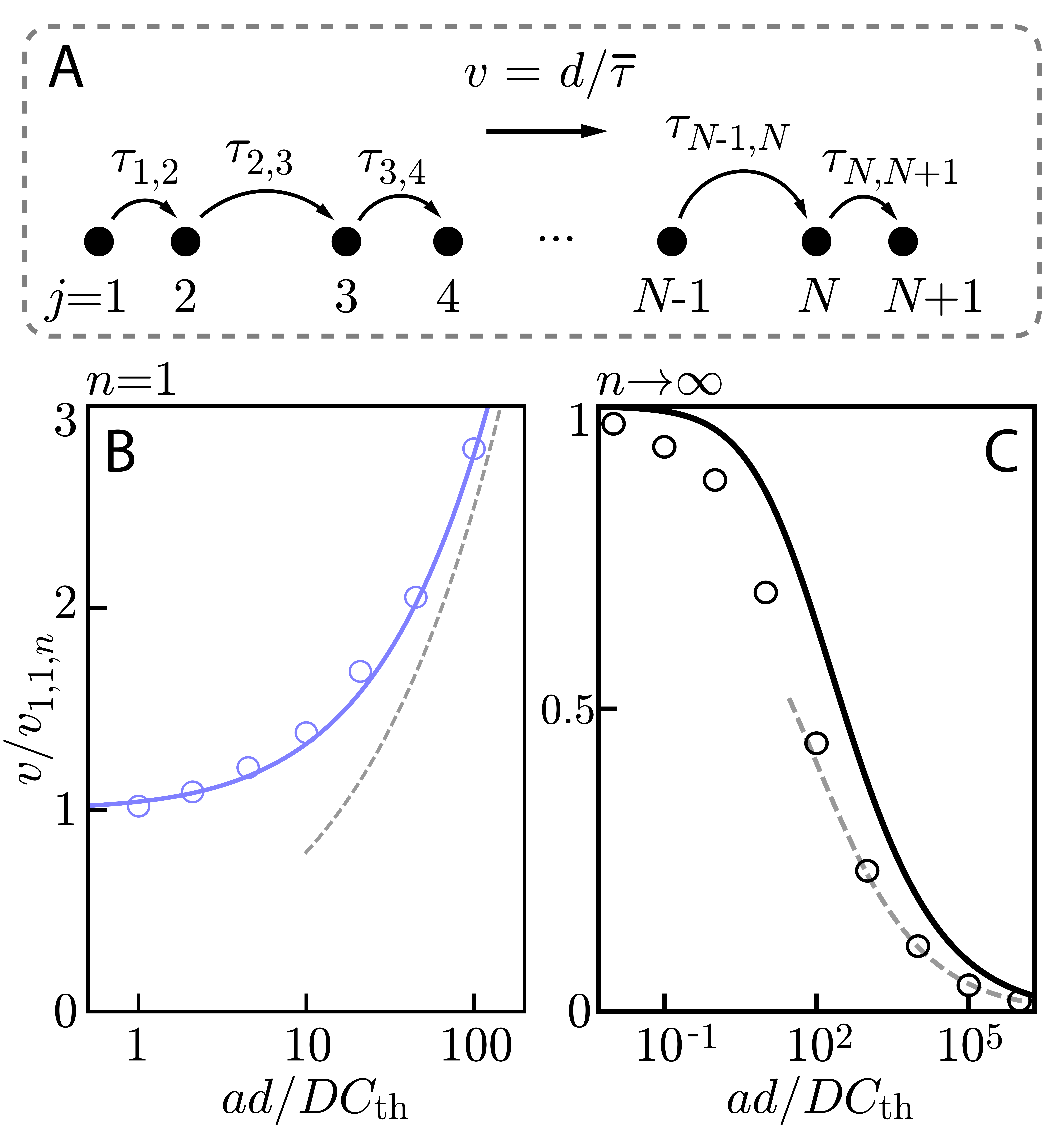}
    \caption{\textbf{Comparison of lattice and disordered wave speeds in 1D.} \textbf{A:} Solution methodology for finding the wave speed propagated by a discrete, disordered system in 1D. The wave speed is determine by the average distance between sources, $d$, divided by the mean hopping time, $\taubar$. In the limit $d\ll D\Cth/a$, $\taubar$ can be solved for analytically. \textbf{B:} Discrete wave speed for $n=1$ divided by the continuum theory wave speed as a function of $ad/D\Cth$. When $v/v_{1,1,n}\approx 1$, the discrete theory is in the continuum limit. Here, we see that the lattice theory (solid line) agrees well with the disordered numerics (circle markers), both of which agree with the nearest-neighbor disordered theory (dashed grey line) in the limit of $ad/D\Cth\gg 1$. \textbf{C:} Same as $\textbf{B}$, but for $n\rightarrow\infty$. Unlike for $n=1$, here we see significant deviation of the disordered numerics (circular markers) and the lattice theory (solid line). In the limit $ad/D\Cth\gg 1$, this discrepancy is well-characterized by our nearest neighbor disordered theory (dashed grey line).}
    \label{fig:2}
\end{figure}

\subsection*{Effects of disorder for \texorpdfstring{$n=1,~n\rightarrow\infty$}{hi}}

Of course, in biology and elsewhere, the sources which propagate diffusive waves are not always found in discrete lattices. As such, we now seek to quantify how similar the dynamics of discrete lattice systems are to the dynamics of disordered relays with discrete sources. Here, we will specialize to the cases of $n=1$ and $n\rightarrow\infty$ as these limits will afford us analytical expressions for the wave speed of disordered systems for $ad/D\Cth\gg 1$. As in the case of lattice systems, the relevant dimensionless parameter for disordered systems is $ad/D\Cth$; this fact can be gleaned through dimensional analysis or by considering the concentration variance generated by a fixed speed diffusive wave propagating through a disordered medium (for details of the latter, see Appendix E).

In a disordered system, a chain of sources in 1D will have some inhomogeneous density of sources (Fig. 2A). The result of this straightforward observation is that the wave front does not propagate smoothly through the system of discrete sources, but rather hops incongruously from source $j$ to source $j+1$ in a time $\tau_{j,j+1}$ (Fig. 2A). One can therefore define the wave speed, $v$, in a disordered 1D system only through an ensemble average in which the average number of sources with the local concentration exceeding threshold at time $t$, $\langle n(t)\rangle$, is related to the wave speed $v$ via the average source separation, $d$:

\begin{equation}
\label{eq:waveSpeedDiscrete}
\langle n(t)\rangle = vt/d
\end{equation}

\noindent where we assume sufficiently large $t$ so that the initial condition is negligible.

In general, it is difficult to fully perform the calculation for $\langle n(t)\rangle$; to see why, consider the case of $n\rightarrow\infty$ and consult Fig. 2A. In order to find the hopping time between source $j$ and source $j+1$, one must know the concentration at source $j+1$ at all times so that one can find the moment at which it exceeds $\Cth$. But this requires knowledge of when sources $j$, $j-1$, etc. turned on (i.e., first exceeded $\Cth$), so that calculating $\tau_{j,j+1}$ requires knowing every previous hopping time.

For all $n$, one can break this endless cycle in the limit $ad/D\Cth\gg 1$ because in that limit, the concentration at every source is generated principally by that source and its nearest neighbor. As a result, the hopping time $\tau_{j,j+1}$ is a function (to be calculated shortly) only of $n$, $D$, $\Cth/a$, and the distance, $d_{j,j+1}$, between source $j$ and source $j+1$. Thus, the hopping time between sources separated by a distance $x$, $\tau(x)$ can be written in terms of the lattice wave speed, $v^\text{lattice}_{1,1,n}$:

\begin{equation}
\label{eq:hoppingTimeXXX}
\tau(x) = x/v^\text{lattice}_{1,1,n}.
\end{equation}

\noindent Then, given some distribution of separations, $p(x)$, with mean $d$, we can calculate $\langle n(t)\rangle$ as $t$ divided by the mean hopping time, $\taubar$:

\begin{multline}
\label{eq:hoppingTimeInt1}
d\gg D\Cth/a:~\langle n(t)\rangle = t/\taubar = \\
t\left(\int_0^\infty dx~\tau(x)p(x)\right)^{-1}.
\end{multline}

\noindent It follows from eq. \eqref{eq:waveSpeedDiscrete} that

\begin{multline}
\label{eq:hoppingTimeInt2}
d\gg D\Cth/a:~v = d/\taubar = \\
d\left(\int_0^\infty dx~\tau(x)p(x)\right)^{-1}.
\end{multline}

\noindent For Poisson-distributed sources with mean separation $d$, $p(x) = e^{-x/d}/d$ and it only remains to calculate $\tau(x)$, a task we shall now undertake for $n=1$ and $n\rightarrow\infty$.

Our task is straightforward for $n=1$ since $v = a/2\Cth$ when $d\gg D\Cth/a$. Thus, $\tau(x) = 2\Cth x/a$ and

\begin{equation}
\label{eq:hoppingTimeInt3}
n=1,~d\gg D\Cth/a:~v = d/\taubar = a/2\Cth
\end{equation}

\noindent by eq. \eqref{eq:hoppingTimeInt2}. Note that this result holds for all $p(x)$ as $d=\langle x\rangle$ and $\taubar = \frac{2\Cth}{a}\int~dx~xp(x) = \frac{2\Cth\langle x\rangle}{a}$. This disordered wave speed is plotted (dashed line) in Fig. 2B along with the lattice theory (solid line) and results from numerical simulations of Poisson-disordered systems (circles). As seen in Fig. 2B, the lattice theory closely corresponds with numerical simulations over several orders of magnitude in $ad/D\Cth$.

For $n\rightarrow\infty$, $\Cth$ can be approximated by truncating eq. \eqref{eq:latticeCthInf} at $j=1$. We then refer to eq. \eqref{eq:cjInf} and note that on a lattice $v = d/\tau$, meaning that

\begin{multline}
\label{eq:hoppingTimeInt4}
n\rightarrow\infty,~d\gg D\Cth/a:~\Cth = \\
\frac{ad}{2D}\left(e^{-d^2/4D\tau}\sqrt{\frac{4D\tau}{\pi d^2}}+\erf\sqrt{\frac{d^2}{4D\tau}}-1\right),
\end{multline}

\noindent a relationship that can be numerically inverted for all $d$, then numerically integrated over to give $v$.

The resulting nearest neighbor theory for $v$ (dashed line) is plotted in Fig. 2C along with the lattice theory (solid line) and results from numerical simulations of Poisson-disordered systems (circles). Here, we see substantial deviation of the lattice theory and the numerical simulations of disordered systems for intermediate and large values of $ad/D\Cth$. However, the disordered system numerics agree very well with our nearest neighbor theory in the limit $ad/D\Cth\gg 1$.

We conclude that in one dimension, the wave-like dynamics of Poisson-disordered and lattice systems correspond very closely for $n=1$ and all values of $ad/D\Cth$. In contrast, with sufficiently large $ad/D\Cth$ and $n\rightarrow\infty$, Poisson-disordered and lattice systems propagate diffusive waves of substantially differing wave speed.

\section*{Discrete model solutions in higher dimensions}

We have fully characterized the diffusive wave dynamics for sources and diffusion in one dimension ($N=M=1$) and now turn our attention to systems of other dimensionalities. Interestingly, we will see that whereas, e.g., the $N=M$ continuum theories all give the same asymptotic wave dynamics, the discrete $N=M=1$ dynamics are distinct from those of $N=M=2,~3$; similarly, the dynamics of $N=M-1=1$ are different from those of $N=M-1=2$.

Unfortunately, models of the form eq. \eqref{eq:modelHillDiscrete} are in general ill-posed for $M\geq 2$. This is due to the fact that the local concentration at a point source diverges in two-or-more dimensions. This results in an infinite self-interaction. Mathematically, one can see this by integrating the Green's function of the diffusion equation for a continuously emitting point source in $M$ dimensions:

\begin{multline}
\label{eq:GFDivergence}
M\geq 2:~c(x=0,t) = \\
a\int_0^td\ttil~\left[4\pi D(t-\ttil)\right]^{-M/2}\rightarrow\infty.
\end{multline}

\noindent Therefore, the self-interaction will diverge for point-like sources with $M\geq 2$ and thus the dynamics of eq. \eqref{eq:modelHillDiscrete} are ill-posed. (The integral in eq. \eqref{eq:GFDivergence} also shows up in a proof of P\'{o}lya's Theorem of recurrent vs. non-recurrent random walks.)

To circumvent this problem, one can consider modeling finite-sized sources or a delay between the emission of a diffusible agent and the measurement of its local concentration, either one of which will eliminate the above divergence. We consider a third alternative: sending $n\rightarrow\infty$. In this limit, the dynamics are well-posed because sources do not interact with their own emissions until they are above the $c=\Cth$ threshold, at which point self-interaction is irrelevant.

We will momentarily perform the task of solving for the wave speeds $v$ within lattice and Poisson-disordered arrangements, but let us first develop intuition for these cases through dimensional analysis. We have committed ourselves to studying models of the form

\begin{equation}
\label{eq:modelHeavisideDiscrete}
\frac{\partial c}{\partial t} = D\nabla^2c+a\Theta[c-\Cth]\sum_j\delta(\mathbf{r}-\mathbf{r}_j),
\end{equation}

\noindent in which we recognize (as we have previously) two independent parameters, $D$ and $\Cth/a$, along with the mean separation between sources $d$.

As above, our task here is to understand how varying $d$ affects the wave speed $v$ for given values of $D$ and $\Cth/a$. For $M=1$, we have seen that we must compare $d$ to the diffusive wave length scale $D/v$ and that doing so reveals the continuum theory to be valid when $d\ll D\Cth/a$.

The same does not hold for $M=2$. Here, $D/v$ is still the length scale of the diffusive wave, but now, with $N=2$, $v\sim \sqrt{aD/d^2\Cth}$ within the continuum theory. Constructing a dimensionless parameter $dv/D \ll 1$ puts no conditions on $d$ and merely requires that $a/D\Cth \ll 1$. This implies that $d$ has nothing to do with the agreement between the discrete and continuum theories. Instead, the agreement is entirely determined by the value of $D\Cth/a$. A similar argument holds for $(N,M)=(1,2)$. This can be understood through dimensional analysis. We are interested in constructing a length scale against which we can compare $d$. The parameters we can use -- by the same arguments as in the $M=1$ case -- are $D$ (units m$^2$/s) and $\Cth/a$ (units m$^2$/s, as distinguished from the units of m/s for $M=1$). Thus, there is no combination of $D$ and $\Cth/a$ which can form a length scale. For fixed values of $D$ and $\Cth/a$, we cannot pack sources closer together and get better agreement with the continuum theory. Nonetheless, our previous intuition that $D$ and $\Cth/a$ ought to be small in order to violate the continuum theory is still valid. Here, the dimensionless parameter which describes the agreement between the discrete and continuum theories is $a/D\Cth$.

The situation is even more extreme for $M=3$. Yet again, we seek to compare $d$ with $D/v$. From our continuum theory results, we know that when $N=3$, $v\sim\sqrt{aD/d^3\Cth}$. Requiring $dv/D\ll 1$ yields a condition that $(dD\Cth/a)^{-1}\ll 1$ -- the agreement between the continuum theory and the discrete theory is \textit{worse} when one decreases $d$ at fixed $D$ and $\Cth/a$. Thus, by cramming sources ever closer together in a 3D diffusive environment, one would weaken the agreement between the discrete system dynamics and those of the continuum. A similar argument holds for $N=2$. We will shortly demonstrate this effect through exact calculations and numerical simulation.

\begin{figure*}[t!]
    \centering
    \includegraphics[width=17.2cm]{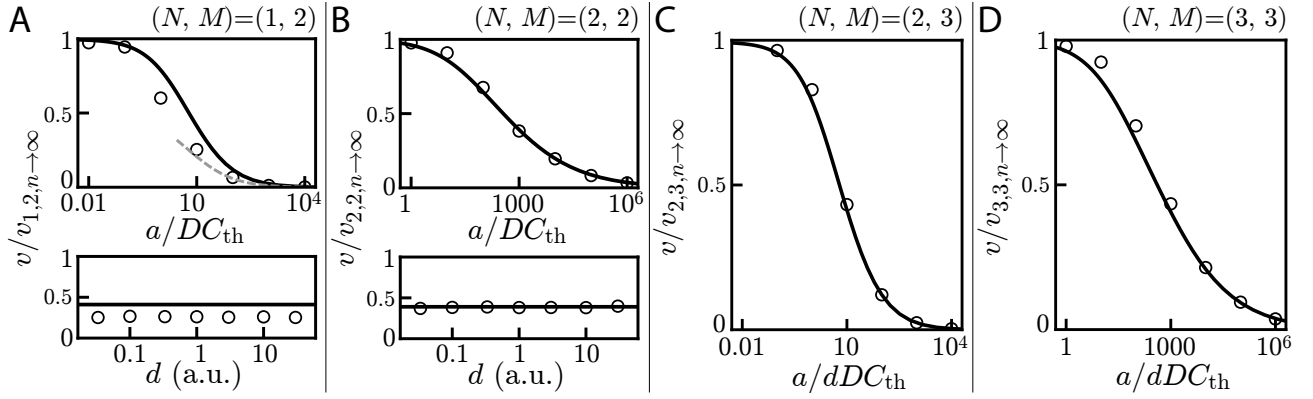}
    \caption{\textbf{Comparison of lattice and Poisson-disordered wave speeds in higher dimensions.} \textbf{A:} Wave speed divided by the continuum theory value with $(N,M) = (1,2)$ for disordered systems (circles) and lattice systems (black line); we plot this quantity for varying $a/D\Cth$ (top) and varying $d$ (bottom). When $v/v_{N,M,n\rightarrow\infty}\approx 1$, a discrete system is in the continuum limit. As predicted by dimensional analysis, $v/v_{1,2,n\rightarrow\infty}$ does not change at all with $d$ (bottom panel), but decreases as $a/D\Cth$ increases (top panel). Results in the bottom panel are shown for $a/D\Cth=10$. The disordered and continuum theories show substantial deviation, which can be understood through a disordered nearest neighbor theory (dashed grey line) at large $a/D\Cth$. \textbf{B:} Same as \textbf{A}, but for $(N,M)=(2,2)$. Here, we see close correspondence between the lattice and disordered system wave speeds. As with $(N,M)=(1,2)$, the correspondence with the continuum theory is determined solely by $a/D\Cth$ (top panel) and does not vary with $d$ (bottom panel). Results in the bottom panel are shown for $a/D\Cth = 1000$. \textbf{C:} Wave speed divided by the continuum theory value as a function of $a/dD\Cth$ with $(N,M)=(2,3)$. Here, we see a close correspondence between the continuum theory (solid line) and the discrete numerical simulations (circles). Unlike the cases of $M=1,2$, decreasing $d$ results in larger deviations from continuum theory. \textbf{D:} Same as \textbf{C} but for $(N,M) = (3,3)$. Here, we see deviations between the continuum theory and numerical simulation for intermediate values of $a/dD\Cth$. Again, we observe that decreasing $d$ results in larger deviations from the continuum theory, as predicted by dimensional analysis.}
    \label{fig:3}
\end{figure*}

\subsection*{Lattice solutions}

Our goal in this section is to systematically find the relationships between $v$, $D$, $d$, and $\Cth/a$ for a set of point sources sitting on a lattice in $N$ dimensions with diffusion in $M$ dimensions. We have already solved the case of $N=M=1$ above (see eqs. \eqref{eq:cjInf} and \eqref{eq:latticeCthInf}). Our methodology for the solutions with other $N,~M$ is largely the same as that already employed, but requires some further elaboration.

To calculate self-consistency relationships analogous to eq. \eqref{eq:latticeCthInf} for $N=2,~3$ we consider a plane wave propagating through a system of sources in $N$-dimensions. With $N=2$, we obtain a picture similar to Fig. 1B, but with sources at $y = 0,~\pm d,\pm 2d, \dots$. At $t=0$, the sources at $x=-jd$ and $y = 0,~\pm d,\pm 2d, \dots$ have been emitting since $t=-jd/v$. If we assume that $c(x=y=t=0) = \Cth$, then we can construct a self-consistency relationship by adding up the contributions from all the point sources left of the origin. For the sake of clarity, we have performed this calculation for $(N, M)=(2,2)$ below; the self-consistency relationship for other system dimensionalities can be found in Appendix F.

\textbf{(N,M)=(2,2):} To find a self-consistency relationship for this dimensionality, we consider the concentration, $c^{(M=2)}_{j,k}$, generated at $x=y=t=0$ by a diffusive point source in two dimensions at $(x,y) = (-jd,kd)$ that has been emitting since $t=-jd/v$. This can be calculated by integrating the diffusion equation Green's function for a 2D environment, a process that yields, with $\Gamma_0[.]$ the incomplete Gamma function,

\begin{multline}
\label{eq:cM2jk}
c^{(M=2)}_{j,k} = -\frac{a}{4\pi D}\int_{-jd/v}^0d\ttil~e^{\frac{j^2d^2+k^2d^2}{4D\ttil}}/\ttil = \\
\Gamma_0\left[\frac{vd}{4Dj}(j^2+k^2)\right]/4\pi D.
\end{multline}

\noindent We may therefore calculate the self-consistency relation for $(N,M)=(2,2)$ by adding up all of the sources:

\begin{equation}
\label{eq:CthN2M2}
(N,M)=(2,2):~\Cth = \sum_{j=1}^\infty\sum_{k=-\infty}^\infty c^{(M=2)}_{j,k}.
\end{equation}

Using eq. \eqref{eq:CthN2M2} and its analogs [eqs. \eqref{eq:CthN1M2}, \eqref{eq:CthN2M3}, and \eqref{eq:CthN3M3}], we have plotted the wave speeds, $v$, for every system dimensionality in Fig. 3 (solid lines). As expected, these relationships reveal that for $M=2$ the discrete-continuum correspondence is entirely independent of $d$ and worsens as $a/D\Cth$ is increased; for $M=3$, the discrete-continuum correspondence worsens as $d$ decreases. Thus, our previously developed intuition and dimensional analysis provide clear intuition for the dynamics of diffusive waves propagated by discrete sources.

\subsection*{Effects of disorder: theory and numerics}

Our goal in this section is to examine the effects of disorder on the diffusive wave propagation in higher dimensional systems. In short, we find that disorder leads to mild deviations from the lattice wave theory developed in the previous section, with the most severe discrepancies for $(N,M)=(1,2)$ (Fig. 3).

To begin, we note that systems with $N\geq 2$ preclude calculation of the wave speed according to the nearest neighbor techniques we developed previously for $(N,M)=(1,1)$. This is due to the fact that in two-and-higher-dimensional spaces, two sources can have the same nearest neighbor. However, one can compute wave speeds in the non-collective limit for $(N,M)=(1,2)$ and can numerically examine the effects of disorder in all system dimensionalities -- tasks we will now undertake.

In the non-collective limit of an $(N,M)=(1,2)$ system, we can arrive at an exact expression for the wave speed by considering the mean hopping time between nearest neighbors. With $v^\text{lattice}_{1,2,n\rightarrow\infty}$ as the wave speed of a lattice theory, the hopping time between two sources separated by a distance $x$ is given by:

\begin{equation}
\label{eq:disordered12}
\tau(x) = x/v^\text{lattice}_{1,2,n\rightarrow\infty}.
\end{equation}

\noindent By our previous dimensional analysis and eq. \eqref{eq:v12Inf}, we know that 

\begin{multline}
\label{eq:disordered12speed}
v^\text{lattice}_{1,2,n\rightarrow\infty} = f(a/D\Cth)~v_{1,2,n\rightarrow\infty} = \\ f(a/D\Cth)\frac{2a}{\pi x\Cth}
\end{multline}

\noindent where $f(a/D\Cth)$ is a function that describes the solid curve plotted in Fig. 3A. Thus, the average hopping time, $\taubar$, is given by

\begin{multline}
\label{eq:taubar12}
\taubar = \int_{0}^\infty dx~p(x)\tau(x) = \\ \frac{\pi\Cth}{2af(a/D\Cth)}\int_0^\infty dx~x^2p(x),
\end{multline}

\noindent which for Poisson-distributed sources with mean separation $d$ gives us

\begin{equation}
\label{eq:vbar12}
v = d/\taubar = f(a/D\Cth)\frac{a}{\pi x\Cth} = v^\text{lattice}_{1,2,n\rightarrow\infty}/2.
\end{equation}

\noindent This relationship agrees with numerical simulations of disordered systems in the large-$a/D\Cth$ limit.

To understand the effect of disorder in the remaining system dimensionalities, we appeal to numerical simulation. As shown in Figure 3, the wave speeds observed in Poisson-disordered systems largely agree with the wave speeds observed in lattice systems of comparable $a/D\Cth$ ($M=2$) or $a/dD\Cth$ ($M=3$).

As expected from dimensional analysis, the effect of varying $d$ in systems with $M=2$ is nil (Figs. 3A/B). This intuition holds for both Poisson-disordered and lattice systems. Meanwhile, increasing $d$ in systems with $M=3$ indeed has the counter-intuitive effect of improving the agreement with the continuum theory (Figs. 3C/D).

In systems with $N>1$, we observe substantial agreement between the lattice theory wave speed and the Poisson-disordered wave speeds obtained in numerical simulation (Figs. 3B-D). This correspondence is not perfect, as can be seen most clearly for values of $1<a/dD\Cth <1000$ with $(N,M)=(3,3)$ (Fig. 3D, we have checked that these are not numerical errors); nonetheless, it is striking and worthwhile to note that disordered dynamics are well-approximated by the more easily calculable properties of lattice theories.

\section*{Discussion}

In this work, we have provided an in depth study of the validity of PDE models for understanding the dynamics of one broad class of biological phenomena. Through parallel studies of the continuum and discrete-source models of diffusion-reaction systems, we have shown that discreteness can affect dynamical properties of these systems in a dimension-dependent and disorder-insensitive manner.

From a physics perspective, many interesting puzzles remain. We have understood the effects of discreteness as a competition between two length scales -- the separation between sources and the inherent length scale of a diffusive wave. Adding additional length and time scales -- the finite size of the source, the finite measurement time of diffusing molecules -- will inevitably result in richer models than those we have considered here. So too would considerations of advection and anomalous diffusion.

From a biological perspective, an even greater zoology of effects remain to be characterized. Understanding the interplay between the external diffusive dynamics and the internal regulatory dynamics of cells will provide a broader model class than we have considered here. Beautiful previous work \cite{muratov2004} in one-dimensional systems has already demonstrated this approach to be an interesting one, as has much work on the chemotactic response of \textit{Dictyostelium discoideum} \cite{kessler1993,palsson1996,noorbakhsh2015}, an organism that uses diffusive waves to guide chemotaxis during the formation of fruiting bodies.

\section*{Acknowledgements}

We thank Chris Rycroft for very helpful discussions of numerical methods. We acknowledge support from the NSF through MRSEC DMR 14-20570 and the Kavli Foundation. P.B.D. is supported by the Paul M. Young Fellowship through the Fannie and John Hertz Foundation. A.A. acknowledges support from NSF CAREER 1752024.

\bibliography{bib}

\newpage

\section*{Appendices}

\subsection*{Appendix A: Self-interaction for \texorpdfstring{$n>1$}{hi}}

Here, we calculate concentration generated at a point source in one diffusive dimension given self-interaction with Hill exponent $n>1$. As mentioned in the main text, we assume a the point-source has been self-interacting since $t\rightarrow-\infty$ and a boundary condition of $c(t=0) = c_0\ll\Cth$. We concern ourselves with the dynamics of $c\ll\Cth$, in which limit the production function is $c^n/(c^n+\Cth^n)\approx c^n/\Cth^n$. We may therefore write down a self-consistency relationship for $c(t)$ by integrating $c(\ttil<t)$ against the Green's function for the diffusion equation:

\begin{equation}
\label{eq:appASelfConsist}
c(t) = \frac{a}{\Cth^n\sqrt{4\pi D}}\int_{-\infty}^td\ttil~c(\ttil)^n/\sqrt{t-\ttil}.
\end{equation}

\noindent With an ansatz of

\begin{equation}
\label{eq:appAnsatz}
c(t) = (\alpha t+\beta)^{-\frac{1}{2(n-1)}},
\end{equation}

\noindent and a variable substitution of $t' = t-\ttil$, we have that

\begin{multline}
\label{eq:appASelfConsist2}
(\alpha t+\beta)^{-\frac{1}{2(n-1)}} = \\ \frac{a}{\Cth^n\sqrt{4\pi D}}\int_{0}^\infty \frac{dt'}{\sqrt{t'}}~\left[\alpha (t-t')+\beta\right]^{-\frac{n}{2(n-1)}},
\end{multline}

\noindent which, performing the integral, gives the following self-consistency relationship (with $\Gamma[.]$ the Gamma function):

\begin{equation}
\label{eq:appAAlpha}
\alpha = -\frac{\Gamma\left(\frac{1}{2(n-1)}\right)^2}{\Gamma\left(\frac{n}{2(n-1)}\right)^2}\frac{a^2}{4D\Cth^{2n}}.
\end{equation}

\noindent Through the initial condition $c(t=0) = c_0$, we ascertain that

\begin{equation}
\label{eq:appABeta}
\beta = c_0^{-2(n-1)}.
\end{equation}

\noindent Thus, the time needed for the concentration to double is roughly:

\begin{equation}
\label{eq:tDoub}
t\sim \left(\Cth/c_0\right)^{2(n-1)}.
\end{equation}

\subsection*{Appendix B: Exponential tails of a diffusive wave}

In this section, we intend to show that diffusive waves propagated among discrete sources on a lattice in one dimension form concentration profiles with exponential or nearly exponential tails. We observe this property generically in numerical simulation and can understand its origin in systems with $n=1$ and $n\rightarrow\infty$.

First, we show exponential tails are a generated by sources with $n\rightarrow\infty$. To do so, we imagine standing at a distance $r+jd$ from a point source that has been emitting at a rate $a$ since $\ttil_j = -jd/v$. The concentration created at a time $t=0$ by source $j$ is thus given by:

\begin{multline}
\label{eq:AppBCj}
c_j(r,t=0)\sim (jd+r)\Bigg[2e^{-\frac{(jd+r)^2v}{4jdD}}\sqrt{\frac{jdD}{(jd+r)^2v}}-\\
\sqrt{\pi}\left(1-\erf\sqrt{\frac{(jd+r)^2v}{4jdD}}\right)\Bigg].
\end{multline}

\noindent To find the concentration at $r$, we must add up the contributions from all $j$,

\begin{equation}
\label{eq:AppBCj3}
c(r) = \sum_{j}c_j,
\end{equation}

\noindent and in the limit of $r\gg D/v,~d$ -- i.e., by looking sufficiently far away from the most recently emitting source -- we may view the sum over $j$ as an integral, in which case:

\begin{equation}
\label{eq:AppBCj4}
c(r) \approx \int dj~c_j(r,t=0) \sim e^{-rv/D}
\end{equation}

\noindent Thus, for $n\rightarrow\infty$, we can expect exponential tails.

The situation is simpler for $n=1$. We have shown in the main text that the concentration at point sources with $n=1$ grows exponentially in time with rate $\gamma v$. Thus, the concentration generated at distance $r$ and time $t=0$ from a point source that has been growing exponentially for a time $T\gg \gamma,r^2/D$ is given by:

\begin{multline}
\label{eq:AppBbleh}
c(r,t=0) \sim \int_{-T}^0d\ttil e^{r^2/4D\ttil}e^{\gamma v\ttil}/\sqrt{-\ttil} \sim \\
\int_{-\infty}^0d\ttil e^{r^2/4D\ttil}e^{\gamma v\ttil}/\sqrt{-\ttil} \sim e^{-r\sqrt{\gamma v/D}}.
\end{multline}

Thus, we can see that exponential tails of concentration profiles are a generic feature of diffusive waves.

\subsection*{Appendix C: Numerics details}

To simulate the dynamics of models of the form eq. \eqref{eq:modelHillDiscrete1D}, we wrote a differential equation solver that uses cubic interpolants to efficiently compute the concentrations at all point sources. Here, we briefly sketch the outlines of this semi-analytic method. The rate of emission, $A_j(t)$, at point source $j$ is given by the local concentration, $c(x_j,t) = c_j(t)$, at that point source:

\begin{equation}
\label{eq:AppC1}
A_j(t) = ac_j^n/(c_j^n+\Cth^n),
\end{equation}

\noindent so therefore

\begin{equation}
\label{eq:AppC2}
dA_j(t)/dt = \frac{an}{\Cth}\left(1-A_j\right)^{1+1/n}A_j^{1-1/n}\frac{dc_j}{dt}.
\end{equation}

\noindent At any given point source, we may calculate the concentration at time $t$ by adding up the contributions from all sources over all prior times according to the diffusion equation Green's function, $G_{ij}(t,\ttil)$:

\begin{multline}
\label{eq:AppC3}
c_j(t) = \sum_i\int_{-\infty}^t d\ttil~A_i(\ttil)\frac{e^{-(x_j-x_i)^2/4D(t-\ttil)}}{\sqrt{4\pi D(t-\ttil)}} = \\
\sum_i\int_{-\infty}^t d\ttil~A_i(\ttil)G_{ij}(t,\ttil)
\end{multline}

\noindent from which one may also calculate $dc_j/dt$ and thus $dA_j/dt$ via:

\begin{multline}
\label{eq:AppC4}
\frac{dc_j(t)}{dt} = \\
\sum_i\left[ A_i(t)G_{ij}(t,\ttil\rightarrow t)+\int_{-\infty}^td\ttil~A_i(\ttil)\frac{dG_{ij}(t,\ttil)}{dt}\right].
\end{multline}

\noindent The activity of solving for $A_j(t)$ then comes down to solving coupled ODEs defined by eq. \eqref{eq:AppC4}. We will do so numerically using cubic interpolants.

Given two time points, $\ttil$ and $\ttil+d\ttil$, the activities $A_j(\ttil)$ and $A_j(\ttil+d\ttil)$ along with their time derivatives $dA_j(\ttil)/dt$ and $dA_j(\ttil+d\ttil)/dt$ allow for the construction of a unique cubic interpolant of the activity between $\ttil$ and $\ttil+d\ttil$. One may then calculate $dc_j/dt$ by performing the integral in eq. \eqref{eq:AppC4} using a cubic interpolant approximation of $A_i(t)$; cubic interpolant has the benefit of having an exact (though algebraically clumsy) analytical expression when integrated against $dG_{ij}(t,\ttil)/dt$.

Thusly, and in combination with standard fourth-order Runge-Kutta methods, we solve for the dynamics of our PDE system, eq. \eqref{eq:AppC4}. An explicit implementation can be found at \url{github.com/pdieterle/discreteDiffWaves}.

\subsection*{Appendix D: Fisher-like waves with \texorpdfstring{$(N,M,n) = (1,2,1)$}{hi}}

Here, we concern ourselves with solving for the wave dynamics of a continuum of sources at density $\rho$ in one dimension ($N=1$) with diffusion in two dimensions ($M=2$) propagating Fisher-like waves ($n=1$). Recall that the continuum dynamics here are given by eq. \eqref{eq:modelHill1D2DODE}, which admits no straightforward solution that we are aware of.

To ascertain the continuum theory dynamics we take a different path using Green's function methods. These techniques are almost identical to the case of $(N,M,n) = (1,1,1)$ discussed in the main text; at a distance $x\gg D/v$ well beyond the wave front, we presume the concentration has a functional form $c(x,t) = c_0 e^{-\gamma x+\gamma vt}\ll \Cth$ (Fig. 1B). Integrating the production against the Green's function gives us a self-consistency relationship of

\begin{multline}
\label{eq:AppD1}
c(x,t) = c_0e^{-\gamma x+\gamma vt} = \\
\frac{a\rho c_0}{2\pi D\Cth}\int_{-\infty}^td\ttil\int_{-\infty}^\infty d\xtil~\frac{e^{-\gamma x+\gamma v\ttil}e^{-(x-\xtil)^2/4D(t-\ttil)}}{t-\ttil}.
\end{multline}

\noindent For $t=0$, we may perform these integrals directly and ascertain that

\begin{equation}
\label{eq:AppD2}
1 = \frac{a}{d\Cth\sqrt{\gamma D(v-\gamma D)}}\implies v = D\gamma+\frac{a^2}{\Cth^2d^2D\gamma},
\end{equation}

\noindent which has a minimum wave speed of $v = 2a\rho/\Cth$, a factor of $\pi$ times larger than the $(N,M,n) = (1,2,n\rightarrow\infty)$ wave speed, in line with previous numerical results \cite{dieterle2020}.

\subsection*{Appendix E: Threshold concentration variance}

We can gain additional intuition for the wave dynamics of discrete, disordered sources by considering the ensemble average (e.g., across experiments) of the concentration variance generated by sources subject to activation by a diffusive wave that travels with constant speed $v$. To see this, we imagine dispersing $n$ sources in one dimension according to a Poisson process. With a uniform density $1/d$ of sources in the region $0\leq x\leq nd$, the probability distribution for any given source is $p(0\leq x\leq nd) = 1/nd$. We assume a traveling wave propagates leftward at a constant speed $v$ through these discrete sources, and that when the wave hits a given source the source begins emitting at a rate $a$. In such a case, the concentration created at $t=x=0$ by a source at $\xtil$ is, for diffusion in 1D, 

\begin{equation}
\label{eq:AppE1}
c(\xtil) = \frac{a\xtil}{2D}\left(e^{-v\xtil/4D}\sqrt{\frac{4D}{\pi\xtil v}}+\erf\sqrt{\frac{v\xtil}{4D}}-1\right).
\end{equation}

\noindent If we consider a source at $\xtil_i$, then its contribution at $x=t=0$ will be $c(\xtil_i) = c_i$. The mean value of $c_i$ can be had by integrating over $0\leq \xtil_i\leq nd$, and the expected concentration created by all $n$ sources is $n$ times this average value. The mean concentration at $x=t=0$ is therefore, as $n\rightarrow\infty$,

\begin{equation}
\label{eq:AppE2}
\langle c(x=t=0)\rangle = \lim_{n\rightarrow\infty}n\int_{0}^{nd}d\xtil~p(\xtil)c(\xtil) = aD/dv^2.
\end{equation}

This is nothing more than the continuous theory value of $\Cth$. To show that the disordered system deviates significantly from the continuum theory for large $vd/D$, we next consider the variance of the concentration at $t=x=0$. As $\langle c^2\rangle$ is simply $n$ times the expectation of $c^2$ generated by a single source, we have

\begin{equation}
\label{eq:AppE3}
\langle c^2(x=t=0)\rangle = \lim_{n\rightarrow\infty}n\int_{0}^{nd}d\xtil~p(\xtil)c^2(\xtil).
\end{equation}

\noindent In the limit $n\rightarrow\infty$, this expression becomes

\begin{equation}
\label{eq:AppE4}
\lim_{n\rightarrow\infty}\langle c^2(t=x=0)\rangle = \frac{2a^2D(10-3\pi)}{3\pi dv^3}+\left(aD/dv^2\right)^2
\end{equation}

\noindent whence the square of the coefficient of variation is

\begin{equation}
\label{eq:AppE5}
\sigma_c^2/\langle c\rangle^2 = (\langle c^2\rangle-\langle c\rangle^2)/\langle c\rangle^2 = \frac{2(10-3\pi)}{3\pi}\frac{dv}{D}
\end{equation}

\noindent which tends to zero as $vd/D\rightarrow 0$ -- the continuous theory limit.

\subsection*{Appendix F: Self-consistency relationships}

In this appendix, we construct the self-consistency relationships governing the wave speeds in systems of dimensionality $(N,M)=(1,2),~(2,3),~\&~(3,3)$ with $n\rightarrow\infty$.

\textbf{(N,M)=(1,2):} We recall that the concentration created at $(x,y)=(0,0)$ by a wave-activated point source at $(x,y)=(-jd,kd)$ with diffusion in $M=2$ dimensions is given by $c_{j,k}^{(M=2)}$ (see eq. \eqref{eq:cM2jk} for precise form). By summing up the $k=0$ terms, we may construct the self-consistency relationship for a one-dimensional lattice of point sources with diffusion in 2D:

\begin{equation}
\label{eq:CthN1M2}
(N,M)=(1,2):~\Cth = 2\sum_{j=1}^\infty c^{(M=2)}_{j,k=0}.
\end{equation}

\noindent Here, the factor of two is needed as we consider a semi-infinite diffusive environment.

\textbf{(N,M)=(2,3):} Following the logic above and in the main text, we consider the concentration, $c^{(M=3)}_{j,k,l}$, generated at $x=y=t=0$ by a diffusive point source in three dimensions at $(x,y,z) = (-jd,kd,ld)$ that has been emitting since $t=-jd/v$. This too can be calculated by integrating the diffusion equation Green's function, a process that yields:

\begin{multline}
\label{eq:cM3jkl}
c^{(M=3)}_{j,k,l} = \frac{a}{\left(4\pi D\right)^{3/2}}\int_{-jd/v}^0d\ttil~e^{\frac{d^2(j^2+k^2+l^2)}{4D\ttil}}(-\ttil)^{-3/2} = \\
\frac{a}{4\pi dD}\frac{1-\erf\sqrt{\frac{vd(j^2+k^2+l^2)}{4jD}}}{\sqrt{j^2+k^2+l^2}}.
\end{multline}

\noindent Similar to the case of $(N,M)=(1,2)$, we may calculate the self-consistency relationship by adding up all the $l=0$ terms:

\begin{equation}
\label{eq:CthN2M3}
(N,M)=(2,3):~\Cth = 2\sum_{j=1}^\infty\sum_{k=-\infty}^\infty c^{(M=3)}_{j,k,l=0}
\end{equation}

\noindent where the factor of two above is used because we again consider a semi-infinite environment.

\textbf{(N,M)=(3,3):} Lastly, we may construct the self-consistency relationship for a lattice of point sources in three dimensions:

\begin{equation}
\label{eq:CthN3M3}
(N,M)=(3,3):~\Cth = \sum_{j=1}^\infty\sum_{k,l=-\infty}^\infty c^{(M=3)}_{j,k,l}.
\end{equation}

\end{document}